\documentclass[aps,pre,showpacs,twocolumn, showemail]{revtex4}
\usepackage{bbm}
\usepackage{mathrsfs}
\usepackage{epsfig,psfrag}
\usepackage{amsmath,amsfonts,amssymb}
\usepackage[usenames]{color}
\usepackage[dvips, bookmarks, colorlinks=true, plainpages = false, citecolor = blue, linkcolor = blue, urlcolor = blue, filecolor = blue]{hyperref}
\def\be{\begin{equation}}
\def\ee{\end{equation}}
\def\bea{\begin{eqnarray}}
\def\eea{\end{eqnarray}}

\begin{document}

\preprint{draft}

\title{Direct Evidence for Conformal Invariance of Avalanche Frontier in Sandpile Models}
\author{A. A. Saberi }\email{a$_$saberi@ipm.ir}
\address
{School of Physics, Institute for Studies in Theoretical Physics and
Mathematics, 19395-5531 Tehran, Iran}

\author{S. Moghimi-Araghi , H. Dashti-Naserabadi and, S. Rouhani }

\address
{Department of Physics, Sharif University of Technology, P.O. Box
11155-9161, Tehran, Iran }
\date{\today}

\begin{abstract}
Appreciation of Stochastic Loewner evolution (SLE$_\kappa$), as a
powerful tool to check for conformal invariant properties of
geometrical features of critical systems has been rising. In this
paper we use this method to check conformal invariance in sandpile
models. Avalanche frontiers in Abelian sandpile model (ASM) are
numerically shown to be conformally invariant and can be described
by SLE with diffusivity $\kappa=2$. This value is the same as value
obtained for loop erased random walks (LERW). The fractal dimension
and Schramm's formula for left passage probability also suggest the
same result. We also check the same properties for Zhang's sandpile
model.
\end{abstract}


\maketitle

\section{Introduction}

The concept of self organized criticality (SOC) was first introduced
by Bak, Tang and Wiesenfeld \cite{BTW} through invention of sandpile
models. These models are still the simplest examples of the class of
models which show self-organized criticality. A definitive step in
analyzing sandpile models was taken in \cite{Dhar}, in which Dhar
introduced a generalization of BTW model. This generalized model was
called Abelian Sandpile Model (ASM), because of the presence of an
Abelian group governing its dynamics. Many different aspects of the
model have been considered, for a good review see \cite{DharRev}. It
was shown that the model could be mapped to spanning
trees\cite{Majumdar1} and is related to $c=-2$ conformal field
theory \cite{Mahieu Ruelle,MRR}.

There is also another non-Abelian sandpile model introduced by Zhang
\cite{Zhang}, which is a continuous version of ASM. Although they
have different microscopic details but it is expected they are in a
same universality class; there has been found numerical evidence for
it \cite{Diaz, Lubeck}.

ASM has been shown to have relation with loop erased random walk
(LERW) \cite{Majumdar0}. The loop erased random walk was proposed by
Lawler \cite{Lawler}. Such a walk is produced by erasing loops in an
ordinary random walk as soon as they are formed. It turns out that
the distribution of the LERW is related to the solution of the
discrete Laplacian \cite{ICTP} with appropriate boundary conditions.
It is also related to the Laplacian random walk \cite{Lap1, Lap2}.
The connection between LERW and ASM arises in the following way
\cite{Majumdar0}: starting from a random walk one can produce a tree
from it called backward tree. Then one can show that the chemical
path on this tree is equivalent with the LERW obtained from the
original random walk. Thus statistical properties of chemical path
on spanning trees and LERW's are the same. Using this
identification, some analytical and numerical results have been
developed. In \cite{Priezzhev} the upper critical dimension of the
ASM was determined and in \cite{KLGP} the above result was confirmed
numerically.

Soon afterwards it was realized that LERW belongs to a family of
conformally invariant curves called Schramm-Loewner evolution,
SLE$_\kappa$, with diffusitivity constant $\kappa=2$ \cite{Schramm0,
Schramm1}. In this paper we show that LERW can be appeared in some
geometrical features of sandpile models generated by their dynamics.
In contrast with the previous results, we do not consider the
chemical path of the spanning trees, but consider the curve
separating the toppleled and untoppled sites i.e. the avalanche
frontier.

This paper is organized as follows. In section 2 we give some
background on the ASM and its properties. Also we introduce Zhang
sandpile model very briefly. Section 3 is devoted to  the definition
and some references on the SLE.  Finally in section 4 we present the
numerical algorithm its results and discussion.

\section{Sandpile model}\label{ASM}

We consider the Abelian Sandpile Model defined on a two dimensional
square lattice $L\times L$. On each site $i$ the height variable
$h_i$ is assigned, taking its value from the set $\{1,2,3,4\}$. This
variable represents the number of sand grains in the site $i$. This
means that a configuration of the sandpile is given by a set of
values $\{h_i\}$.

The dynamics of the system is relatively simple. At each time step,
a grain of sand is added to a random site, $i$. Then site is checked
for stability, that is if its height is more than $4$, it becomes
unstable and topples: it loses $4$ grains of sands, each of them is
transferred to one of the four neighbors of the original site. It is
common to write $h_j \rightarrow h_j - \Delta_{ij}$ for all $j$ with
$\Delta$ being discrete Laplacian. As a result of a toppling, the
neighboring sites may become unstable and topple and a chain of
topplings may happen in the system. If a boundary site topples, one
(or two) grains of sand may leave the system, depending on the
imposed boundary condition taken. The chain of topplings continue
until the system becomes stable, i.e., all the height variables
become less than or equal to four. Thus in each time step, the
dynamics takes the system from a stable configuration $C_m$ to
another stable configuration $C_{m+1}$. The relaxation process is
well defined: it always stops because sand can leave the system at
the boundaries, and produces the same result independent of the
order in which the topplings are performed which is because of the
Abelian property.

Under this dynamics the system reaches a well-defined steady state.
All the stable configurations fall apart into two subsets: the
transient states that do not occur in the steady state and the
recurrent states that all occur with the same probability. It has
been shown that the total number of recurrent states is $\det
\Delta$ \cite{Dhar}. The criterion that decides whether a
configuration is recurrent or not is not a local one. There are some
specific clusters, called forbidden subconfigurations (FSC's) that
if any of them is found in a stable configuration, it would be a
transient configuration. The simplest FSC is a cluster of two
adjacent height-one sites. In general an FSC is a height
configuration over a subset of sites, such that for any of the sites
in this subset, the number of its neighbors within the same subset,
is greater than or equal to its height. Such subsets could be as
large as the whole system, thus in general you can not decide easily
if a configuration is recurrent or not.

An interesting question would be what is the probability of finding
a site with height $h$, or what is the probability of finding a
specific cluster of height variables. Even more interesting, is the
joint probabilities of such events. These questions have been
answered for the case of Weakly Allowed Clusters (WACs)
\cite{Majumdar1}. WACs are the clusters that are not FSC, but if you
remove a grain of sand from any of its sites it becomes FSC. The
simplest example is one-site height-one cluster.

The correlation functions of all such clusters obey a power law with
the same exponent; all the clusters have scaling exponent equal to
two. From point of view of critical systems, one expects that in the
scaling limit ASM should be expressed via a field theory. There have
been found many indications that a specific conformal field theory
called the $c=-2$ theory is related to ASM. First of all a
connection between ASM and spanning trees has been found
\cite{Majumdar1}, therefore it should be related to $q\rightarrow 0$
Potts model, which is known to be related to the $c=-2$ theory. Also
the exponents of the WAC fit in this theory. In \cite{Mahieu Ruelle}
the critical and off-critical two- and three-point correlation
functions of $14$ simplest WACs were calculated and using these
results the scaling fields associated with these WACs were obtained.
This result was generalized to arbitrary WAC in \cite{Jeng}. These
identifications were done only by comparing the correlation
function. In \cite{MRR} the fields were derived from an action and
the way the probabilities are calculated in ASM are translated
directly to field theory language to obtain the relevant fields. The
$c=-2$ theory is a logarithmic theory \cite{c=-2} and it contains
some fields that have logarithmic terms in their correlation
functions. Such fields are related to one-site clusters with height
more than one \cite{JPR}, though still a direct way to show it, is
missing. The action of $c=-2$ is $S\sim \int \partial
\bar{\theta}\bar{\partial}\theta$ where $\theta$ and $\bar{\theta}$
are Grassmannian variables. It is easy to see why the action is
related to ASM, just note that the number of recurrent
configurations is $\det\Delta$ and all occur with the same
probability. So the partition function of the system is
$\det\Delta$. This determinant could be written in terms of
integrating over Grasmannian variables which leads to the above
action in the scaling limit.

Interestingly, it was observed in \cite{ICTP} that the probability
distribution of LERW may be written in terms of a Grasmannian path
integral, reinforcing the connection between LERW and ASM.

Different properties characterizing an avalanche is the other
subject usually investigated in ASM. We call the total number of
topplings the size of avalanche and denote it by $s$. The number of
distinct lattice sites toppled is denoted by $d$ which is clearly
less than or equal to the size of avalanche. This variable shows the
area of the system which is affected by the avalanche. The duration
$t$ of an avalanche is the number of update sweeps needed until all
sites are stable again. The other characteristic is the linear size
of an avalanche which is measured via the radius of gyration of the
avalanche cluster and is denoted by $R$. In the critical steady
state the corresponding probability distributions obey power-law
behavior \be \label{Palpha} P_\alpha(\alpha) \sim
\alpha^{-\tau_\alpha}, \ee
 where $\alpha$ can be $s$, $d$, $t$ or
$R$. These exponents are calculated numerically \cite{Manna,Olami},
also using specific assumptions some (different) analytic results
have been obtained \cite{analytic}. The exponents are not
independent, as an example because the region that the sites topple
is a compact one and does not have holes in it, the area $s$ of the
region should be proportional to $R^2$ statistically. This induces
the relation $\tau_r=2\tau_s+1$ between the exponents.

Other versions of sandpile models have been considered
\cite{Zhang,Manna1}. In \cite{Zhang}, Zhang introduced a model in
which the height variables were continuous and are called energy. At
any time step a random amount of energy is added to a random site.
If the energy of the site becomes more than a specific amount,
called threshold, it becomes active and topples: it loses all its
energy, which is equally distributed among its nearest neighbors. In
his original paper, Zhang observes, based on results of numerical
simulation, that for large lattices, in the stationary state the
energy variables tend to concentrate around discrete values of
energy; he calls this the emergence of energy 'quasi-units'. Then,
he argues that in the thermodynamic limit, the stationary dynamics
should behave as in the discrete ASM. Zhang model dose not have the
Abelian property, therefore little analytic results is at hand.
However the numerical simulations show that it exhibits finite size
scaling property Eq. \ref{Palpha} \cite{Lubeck, Lubeck1}.

\begin{figure}\begin{center}
\epsfxsize=8. truecm\center\epsfbox{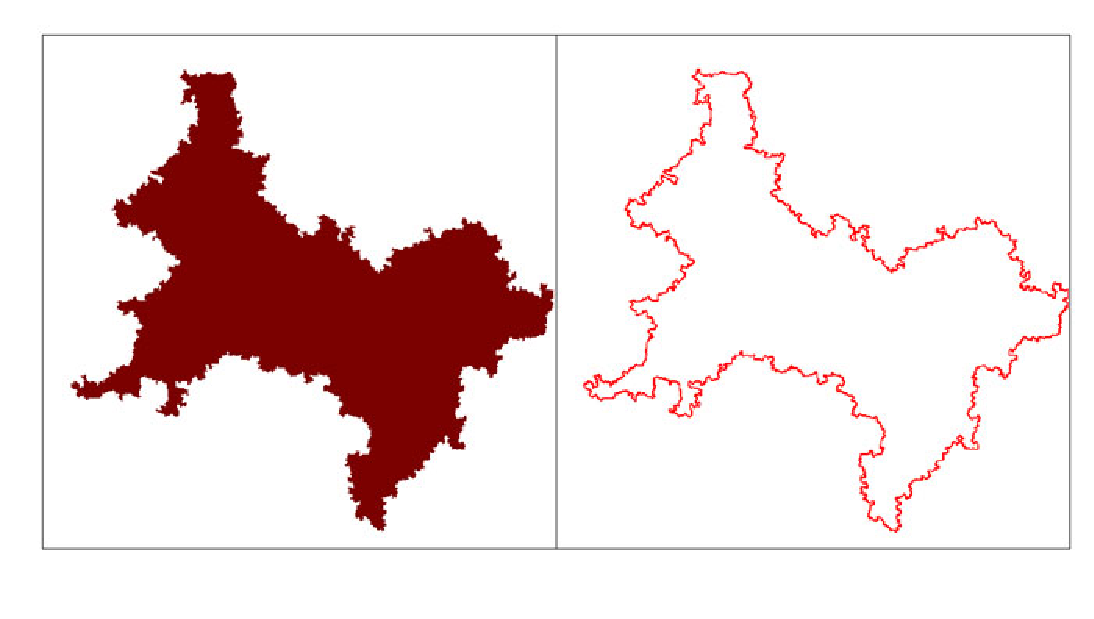}
\caption{\label{Fig1}(color online). An avalanche cluster (left)
consisting all sites that have toppled at least once, and, its
frontier (right).}\end{center}
\end{figure}

These scaling relations imply that there should be some related
geometric structures in the avalanches. We consider avalanche
clusters in the steady state in which all sites have experienced
toppling at least once. Then, in the following sections using theory
of SLE, we investigate the statistics of the avalanche boundaries
(see Fig. \ref{Fig1}).

\section{Stochastic Loewner Evolution}\label{SLE}

Critical behavior of various systems can be coded in the behavior of
their geometrical features.  In two dimensions, the criticality
shows itself in the statistics of interfaces e.g. domain walls. The
domain walls are some non-intersecting curves which directly reflect
the status of the system in question.  For example, consider one of
the prototype lattice models which can be interpreted in terms of
random non-intersecting paths, Ising model, which we consider it in
the physical domain i.e. upper half plane $\mathbb{H}$. To impose an
interface growing from zero on the real line to infinity, a fixed
boundary condition can be considered in which all spins in the right
and left sides of the origin are up and down respectively. At zero
temperature the interface is a straight line and increasing the
temperature leads the interface to a random non intersecting curve.
In the $1920s$, it has been shown by Loewner \cite{Loewner} that any
such curve in the plane which does not cross itself can be created,
in the continuum limit, by a dynamical process called Loewner
evolution with a suitable continuous driving function $\xi_t$ as

\be \label{Loewner} \frac{\partial g_t(z)}{\partial
t}=\frac{2}{g_t(z)-\xi_t}, \ee

Where, if we consider the hall $K_t$, the union of the curve and the
set of points which can not be reached from infinity without
intersecting the curve, then $g_t(z)$ is an analytic function which
maps $\mathbb{H}\setminus K_t$ into the $\mathbb{H}$ itself.\\
For the mentioned Ising model, at zero temperature the interface can
be described in the continuum limit by Loewner evolution with a
specific constant driving function. At higher temperatures less than
critical temperature $T_c$, the driving function might be a
complicated random function. At $T=T_c$,  the system and the
interfaces as well,  are conformally invariant (in an appropriate
sense) i.e. they are invariant under local scale transformations.
Schramm has shown \cite{Schramm0} that the consequences of conformal
invariance for a set of random curves are such that the driving
function in the Loewner evolution should be proportional to a
standard Brownian motion $B_t$ (which is known as stochastic-Schramm
Loewner evolution or SLE$_\kappa$). Therefore
$\xi_t=\sqrt{\kappa}B_t$ so that $\langle \xi_t\rangle=0$ and
$\langle (\xi_t-\xi_s)^2\rangle=\kappa |t-s|$ (for more precise
mathematical definitions and theorems see the review articles
\cite{Review} and references therein).

The diffusivity $\kappa$ classifies different universality classes
and is related to the fractal dimension of the curves $D_f$ as
\be\label{Df} D_f=1+\kappa/8. \ee

After invention of SLE, many of its properties and applications have
been appeared by both mathematicians and physicists. Its connection
with conformal field theory has also been made explicit in a series
of papers by Bauer and Bernard \cite{bernard0}. It has been also
appeared in various physical subjects such as two dimensional
turbulence \cite{bernard1, bernard2}, spin glasses \cite{spin
glass}, nodal lines of random wave Functions \cite{Nodal},
experimental deposited $WO_3$ surface \cite{WO3} and also in two
dimensional Kardar-Parisi-Zhang surface \cite{KPZ}. The connection
between SLE and some lattice models in the scaling limit is also
proven or conjectured today. For example, two dimensional loop
erased random walk (LERW) is a random curve, whose continuum limit
is proven to be an SLE$_2$ \cite{Schramm0}. Self avoiding random
walk (SAW) \cite{Tom-Kennedy} and cluster boundaries in the Ising
model \cite{Smirnov}, are also conjectured to be SLE$_{8/3}$ and
SLE$_3$, in the scaling limit, respectively.

One of the calculations has been made by SLE which will be referred
later, is the probability that the trace of SLE in domain
$\mathbb{H}$, passes to the left of a given point at polar
coordinates $(R, \phi)$. It was studied by Schramm using the theory
of SLE in \cite{Left}. Because of scale invariance, this probability
depends only on $\phi$ and has been shown that \be \label{Left}
P_{\kappa}(\phi)=\frac{1}{2}+\frac{\Gamma\left(\frac{4}{\kappa}\right)}{\sqrt{\pi}\Gamma\left(\frac{8-\kappa}{2\kappa}\right)}
{}_2F_1\left(\frac{1}{2},\frac{4}{\kappa};\frac{3}{2};-\cot^2(\phi)\right)\cot(\phi).
\ee

In the following, we will use these statements to show that the
avalanche frontiers in the both ASM and Zhang's model can be
described by SLE$_2$.

\section{Numerical details: Test for Conformal Invariance}

\begin{figure}\begin{center}
\includegraphics[width=3.7 in, height=3. in]{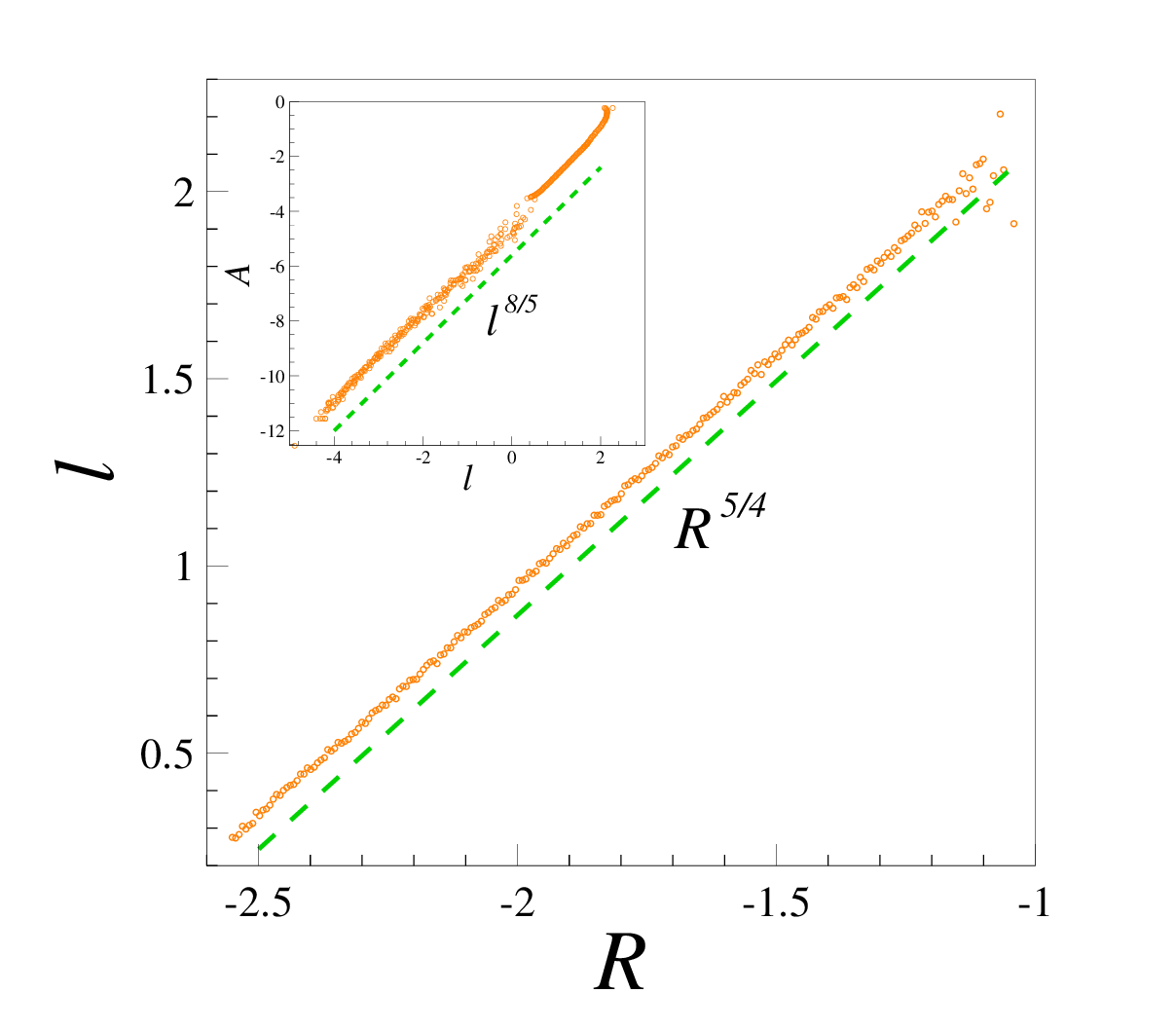}
\caption{\label{Fig3}(color online). Main frame: Log-log plot of The
perimeter of avalanche frontiers (loops) $l$ versus the radius of
gyration $R$, for ASM model simulated on squared lattice with size
of $1024^2$. Inset: Log-log plot of the average area of loops $A$ vs
the length $l$. The dashed lines show the results for
LERW.}\end{center}
\end{figure}

In this section, using the scaling relations and theory of
stochastic Loewner evolution introduced in previous sections, we
show that the conformal field theory which describes the sandpile
models algebraically, can be derived from a quit different approach
i.e., investigation of the statistics and symmetries of some
well-defined geometric features during the sandpile dynamics. To
this end, we consider the avalanche clusters in the steady state
regime during the dynamics: including all sites which topple at
least once at each time step when adding a grain to a random site of
the system makes it unstable (see section \ref{ASM}). Then we get an
ensemble of the boundary of these clusters as suitable candidates to
study their statistics and possible conformal invariance. We compare
our results with similar ones for known models which their relations
with sandpile models is made explicit i.e., LERW.

\begin{figure}\begin{center}
\includegraphics[width=3.5 in, height=3 in]{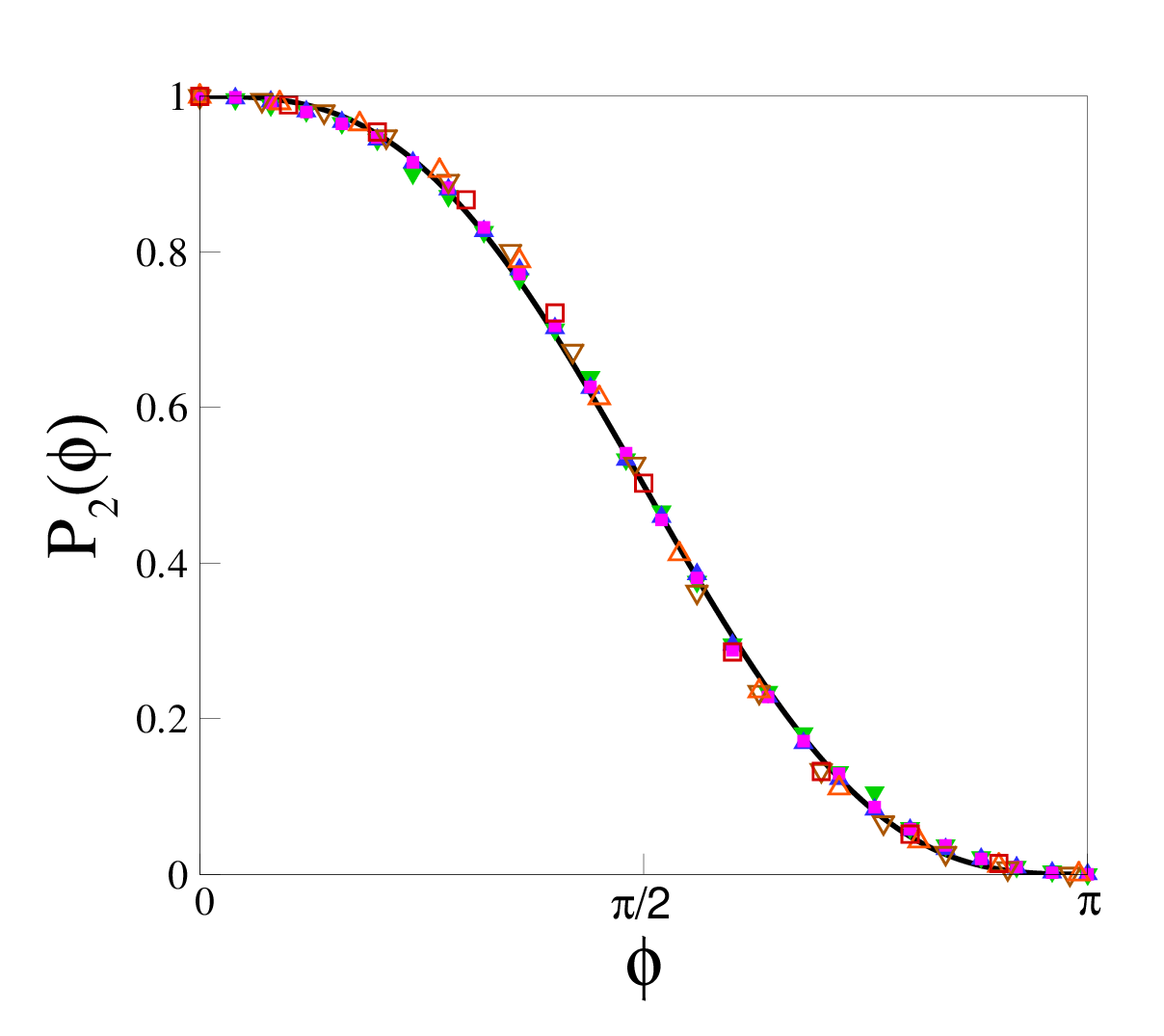}
\caption{\label{Fig4}(color online). The probability that an
avalanche frontier of ASM model (filled symbols) and Zhang model
(open symbols) in domain $\mathbb{H}$, passes to the left of a point
at polar coordinates $(R, \phi)$, for $R=0.05, 0.1$ and $0.2$. The
solid line shows the prediction of SLE for $\kappa=2$ ($P_{2}(\phi)$
in ($\ref{Left}$)). }\end{center}
\end{figure}

To investigate the statistical behavior of the avalanche boundaries
(loops) in the ASM model, we first calculate their fractal dimension
by using the scaling relation between their radius of gyration $R$,
and their perimeter length $l$, i.e., $l\sim R^{D_{f}}$. As shown in
Fig. \ref{Fig3}, the fractal dimension is very close to the one for
LERW which is proven to be $5/4$, (the best fit to the data yields
$D_f=1.24\pm0.02$). \\It is also discussed in \cite{Cardy} that the
mean area of the loops scales with their perimeter length as $A\sim
l^{2/D_{f}}$. The inset of Fig. \ref{Fig3} shows the comparison of
this relation with one calculated for avalanche boundaries in ASM
model. The same results can be obtained
for the avalanche boundaries in Zhang's model which have not been shown here.\\
This fractal dimension $D_f$ is consistent with the fractal
dimension of SLE$_2$ curves in the scaling limit (see Eq. \ref{Df}).
This suggests that the scaling limit of the avalanche frontiers may
be conformally invariant in the same universality class of LERW.

\begin{figure}\begin{center}
\includegraphics[width=3.4 in, height=2.8 in]{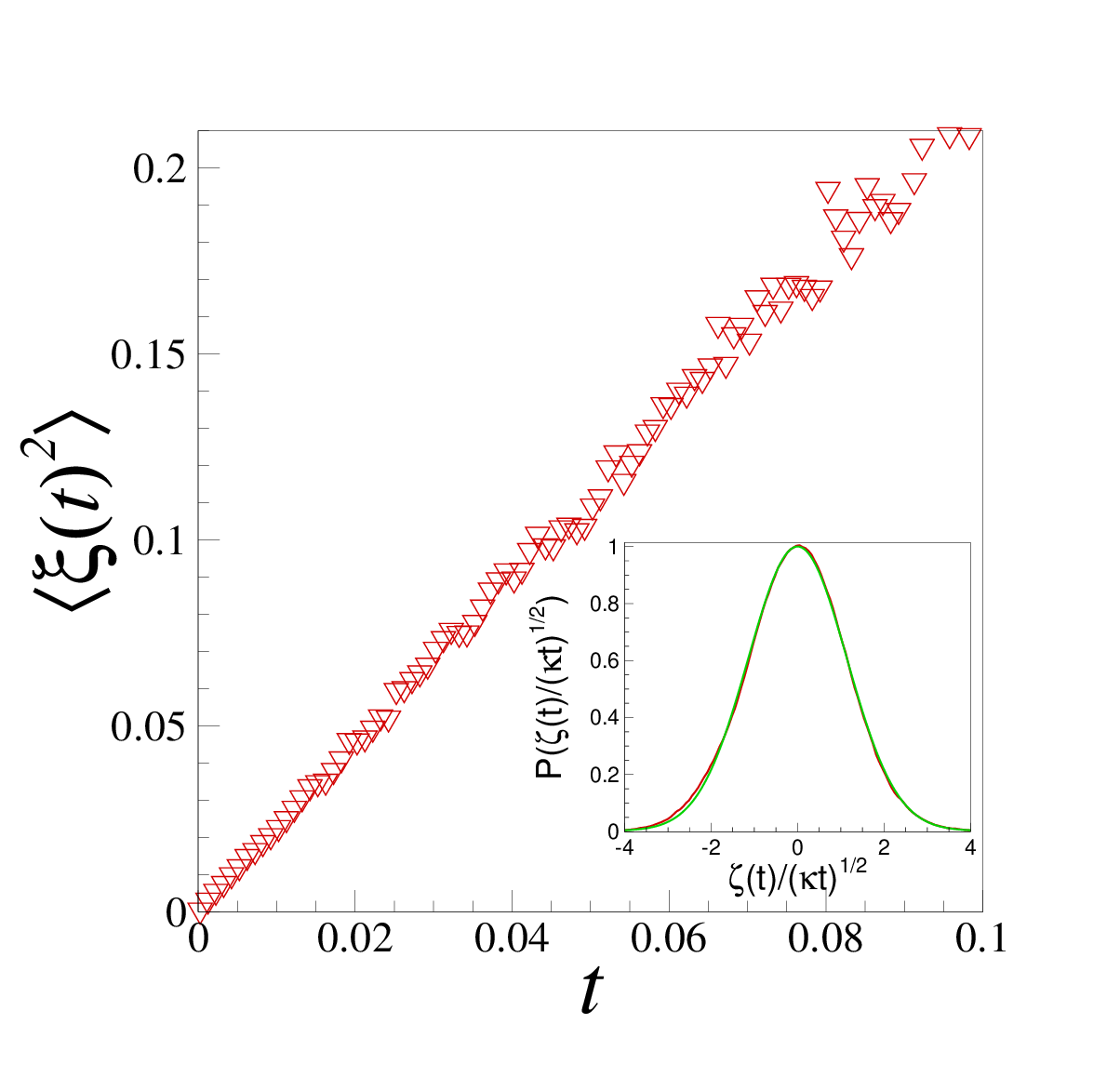}
\narrowtext\caption{\label{Fig5}(color online). Statistics of the
driving function $\xi(t)$ for the avalanche boundaries of ASM model.
Main frame: the linear behaviour of $\langle\xi(t)^2\rangle$ with
the slope $\kappa=2.1\pm0.1$. Inset: the probability distribution
function of the noise $\xi(t)/\sqrt{\kappa t}$ for $0\leq t\leq
0.05$}.\end{center}
\end{figure}

A simple way to check this proposition can be done using Eq.
\ref{Left}. Since in this equation it is supposed that the curves
are in domain $\mathbb{H}$, so we have to be careful about reference
domain. We assume that any avalanche frontier is in the plane, and
then we can consider any arbitrary straight line which crosses the
loop at two points $x_0=0$ and $x_\infty$, as real line. Then we cut
the portion of the curve which is above the real line. To have a
curve starting from origin and tending to infinity, we use the map
$\varphi(z)=x_\infty z/(x_\infty-z)$ for all points of the curve
\footnote{To preserve the reflection symmetry, we take the sign of
$x_\infty$ for each curve, plus or minus with probability
$\frac{1}{2}$. }. Doing so for all frontiers, we would have an
ensemble of such curves and we can check Schramm's formula ( Eq.
$\ref{Left}$) for them.

Fig. $\ref{Fig4}$ shows the result for avalanche frontiers of both
ASM and Zhang's model. The result is most consistent with the
prediction for SLE$_2$ curves.

Now we are in a position to extract the Loewner driving function
$\xi_t$, in Eq. $\ref{Loewner}$, for these avalanche boundaries and
examine whether they are Brownian motion. This is another direct
check which shows the behavior of the curves under local scale
transformations. We use successive conformal maps according to the
algorithm introduced by Bernard \textit{et al.} \cite{bernard2}
based on the approximation that driving function is a piecewise
constant function.\\
The procedure is based on applying the map $G_{t,\xi}=x_\infty
\{\eta
x_\infty(x_\infty-z)+[x_\infty^4(z-\eta)^2+4t(x_\infty-z)^2(x_\infty-\eta)^2]^{1/2}\}
/\{x_\infty^2(x_\infty-z)+[x_\infty^4(z-\eta)^2+4t(x_\infty-z)^2(x_\infty-\eta)^2]^{1/2}\}$
on all the points $z$ of the curve approximated by a sequence of
$\{z_0=0, z_1, \cdot\cdot\cdot, z_N=x_\infty\}$ in the complex
plane, where $\eta=\varphi^{-1}(\xi)$ and again $\varphi(z)=x_\infty
z/(x_\infty-z)$. In which the dimensionless parameter $t$, is used
for parametrization of each curve. At each step, by using the
parameters $\eta_0=\varphi^{-1}(\xi_0)=[\Re z_1 x_\infty-(\Re
z_1)^2-(\Im z_1)^2]/(x_\infty-\Re z_1)$ and $t_1=(\Im
z_1)^2x_\infty^4/\{4[(\Re z_1-x_\infty)^2+(\Im z_1)^2]^2\}$, one
point of the curve $z_0$ is swallowed and the resulting curve is
rearranged by one element shorter. This operation
yields a set containing $N$ numbers of $\xi_{t_k}$ for each curve.\\
Fig. \ref{Fig5} shows analysis of statistics of the ensemble of the
driving functions. Within the statistical errors, it converges to a
Gassian process with the linear behavior of
$\langle\xi(t)^2\rangle$, and the slope $\kappa=2.1\pm0.1$.\\ The
predicted universality class for avalanche frontiers of sandpile
models with diffusivity $\kappa=2$ is consistent with the central
charge of conformal field theory with $c=-2$, given by the relation
$c=(8-3\kappa)(\kappa-6)/2\kappa$, which is supposed to define the
ASM model \cite{Mahieu Ruelle,MRR}.

All these evidences show another example that the theory of SLE can
define (or predict) the conformal field theory which describes the
system.

\section{conclusion}

In this paper, we analyzed the statistics of avalanche frontiers
that appear in the geometrical features of sandpile dynamics. Using
the theory of SLE, we found numerically that the curves are
conformally invariant with the same properties as LERW, with
diffusivity of $\kappa=2$. This relation with LERW which has been
obtained in a quit different way, with respect to the previous
studies, suggests that logarithmic conformal field theory with
central charge $c=-2$, defining the system is in agreement with that
obtained from algebraic approach.

The avalanche front is expected to be an SLE$_2$ from circumstantial
evidence. The ASM model has been argued to be related to $c=-2$
logarithmic conformal field theory which is turn is related to SLE
with $\kappa$ equal to either $2$ or $8$. However as $\kappa = 8$ is
a space filling curve, not a good candidate for the avalanche front
leaving us with $\kappa= 2$. A more definite reasoning, we note that
the way an avalanche is formed one can define a burning algorithm:
at each step, the site $i$ topples if its height $h_i$ is larger
than the number of those of its nearest neighbors which have not
toppled in the previous step. This burning algorithm leads to a tree
that spans the whole area of the avalanche. Hence the avalanche
front is expected to be the dual of the spanning tree thus must have
$\kappa= 2$.

\end{document}